\documentclass[aip,jcp,reprint,superscriptaddress,longbibliography]{revtex4-1} 
\usepackage[utf8]{inputenc}

\usepackage{cmap} 
\usepackage[T1]{fontenc} 
\usepackage{microtype}
\usepackage{amsmath,mathtools,mleftright}
\usepackage{lmodern}
\usepackage{amssymb,bm} 

\providecommand\what\hat 
\providecommand\wtilde\tilde 

\usepackage{siunitx}
\usepackage{graphicx}

\usepackage{booktabs}
\usepackage{dcolumn}
\newcolumntype{d}[1]{D{.}{.}{#1}}

\usepackage{hyperref}
\usepackage[table,usenames,dvipsnames]{xcolor}
\hypersetup{colorlinks=true, linkcolor=BrickRed,
  urlcolor=blue!50!black, citecolor=blue!50!black}

\usepackage{cleveref}

\usepackage{multibib}
\newcites{M}{References}

\renewcommand\epsilon{\varepsilon}
\renewcommand\phi{\varphi}
\renewcommand\theta{\vartheta}
\renewcommand\rho{\varrho}
\renewcommand\vec[1]{{\boldsymbol #1}}

\newcommand\diff{\mathrm{d}}
\newcommand\expect[1]{\left\langle{#1}\right\rangle}
\newcommand\e{\text{e}}
\renewcommand\i{\text{i}}
\renewcommand\geq\geqslant
\renewcommand\leq\leqslant

\DeclareMathOperator\Imag{Im}
\DeclareMathOperator\Real{Re}
\DeclareMathOperator{\Ei}{Ei}  

\newcommand\kB{k_{\text{B}}}
\newcommand\vth{v_\mathrm{th}}
\newcommand\MSD{\mathrm{MSD}}
\newcommand\Y{{\what{\mathcal{Y}}}}
\DeclareTextSymbolDefault{\textdegree}{TS1}

\usepackage[normalem]{ulem}

\begin{document}

\title{Emergence of molecular friction in liquids: bridging between the atomistic and hydrodynamic pictures}

\author{Arthur V. Straube}%
\email{arthur.straube@fu-berlin.de}
\affiliation{Freie Universit{\"a}t Berlin, Department of \rlap{Mathematics and Computer Science, Arnimallee 6, 14195 Berlin, Germany}}
\affiliation{Zuse Institute Berlin, Takustr. 7, 14195 Berlin, Germany}

\author{Bartosz G. Kowalik}%
\affiliation{Freie Universit{\"a}t Berlin, Department of Physics, \rlap{Arnimallee 14, 14195 Berlin, Germany}}

\author{Roland R. Netz}%
\affiliation{Freie Universit{\"a}t Berlin, Department of Physics, \rlap{Arnimallee 14, 14195 Berlin, Germany}}

\author{Felix H{\"o}f{}ling}%
\email{f.hoefling@fu-berlin.de}
\affiliation{Freie Universit{\"a}t Berlin, Department of \rlap{Mathematics and Computer Science, Arnimallee 6, 14195 Berlin, Germany}}
\affiliation{Zuse Institute Berlin, Takustr. 7, 14195 Berlin, Germany}


\begin{abstract}
\bgroup\bfseries 

Friction in liquids arises from conservative forces between molecules and atoms.
Although the hydrodynamics at the nanoscale is 
subject of intense research
and despite the enormous interest in the non-Markovian dynamics of single molecules and solutes,
the onset of friction from the atomistic scale so far could not be demonstrated.
Here, we fill this gap based on frequency-resolved friction data from high-precision simulations of three prototypical liquids, including water.
Combining with rigorous theoretical arguments, we show that friction in liquids emerges abruptly at a characteristic frequency, beyond which viscous liquids appear as non-dissipative, elastic solids;
as a consequence, its origin is non-local in time.
Concomitantly, the molecules experience Brownian forces that display persistent correlations and long-lasting memory.
A critical test of the generalised Stokes--Einstein relation,
mapping the friction of single molecules to the viscoelastic response of the macroscopic sample,
disproves the relation for Newtonian fluids,
but substantiates it exemplarily for water and a moderately supercooled liquid.
The employed approach is suitable to yield novel insights into vitrification mechanisms and the intriguing mechanical properties of soft materials.

\egroup 
\end{abstract}

\maketitle


\begin{figure*}%
  \includegraphics{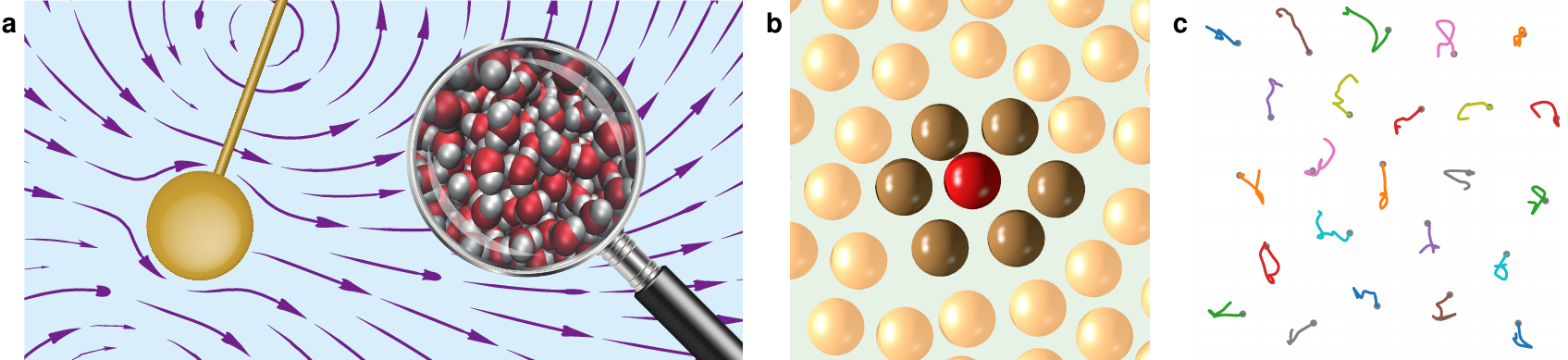}
  \caption[]{
    \textbf{Dynamic friction bridges between the hydrodynamic and atomistic pictures of liquids.}
    Panel~(a): A pendulum that oscillates in a viscous fluid with frequency $\omega$ experiences a dynamic friction $\zeta(\omega)$ as calculated by Stokes (1851) for slow motion. 
    The associated flow pattern (stream lines) leads to hydrodynamic memory of the motion.
    Magnifying the microscopic scale, the fluid consists of molecules that obey Newton's equations, which are non-dissipative and generate smooth trajectories.
    Stokes's result for the pendulum scales down to single molecules, and the function $\zeta(\omega)$ provides the bridge between the frictionless (microscopic) and the hydrodynamic (macroscopic) descriptions.
    ~Panels~(b,c): In liquids, molecules rattle in transient cages formed by their neighbours (brown spheres);
    the corresponding short-time trajectories are smooth, but chaotic curves (panel c).
    The onset of friction is driven by the momentum transfer to the cage, as supported by control simulations of a single particle (red sphere) in a matrix of pinned particles (brown and yellow).
    Illustration for a mono-atomic fluid in two dimensions.
  }
  \label{fig:sketches}
\end{figure*}

Molecular friction is a key ingredient for dynamic processes in fluids: it limits diffusion, governs dissipation, and enables the relaxation towards equilibrium.
In a liquid environment,
the friction experienced by solvated molecules and nanoprobes exhibits a
delayed response to external stimuli, indicating non-Markovian dynamics
\cite{Franosch:N2011,Kheifets:S2014,Berner:NC2018,Daldrop:PRX2017}.
Such memory is found on sub-picosecond up to microsecond scales; it has repercussions on macromolecular transition rates \cite{Guerin:N2016,Sancho:NC2014,Daldrop:PNAS2018}
and is manifest in the visco-elastic behaviour of soft materials
\cite{Sollich:PRL1997,Mizuno:S2007,Winter:PRL2012,Hoefling:RPP2013}.
However, the origin of friction from conservative forces between molecules and atoms remains as one
of the grand challenges of the physics of fluids
\cite{Secchi:N2016,Perakis:NC2018,Bylinskii:S2015,Gaspard:N1998}.

Stokes's friction law, describing the resistance to a steadily dragged immersed sphere of radius $a$, links the friction $\zeta_0$ to the (macroscopic) shear viscosity $\eta_0$, and the relation $\zeta_0 = 6\pi\eta_0 a$ scales down remarkably well to single molecules \cite{Bartsch:PRL2010,Hansen:SimpleLiquids}.
Stokes's hydrodynamic treatment \cite{Stokes:1851} from 1851 was actually more general and addressed slow oscillatory motions in viscous fluids, motivated by inaccuracies of pendulum clocks caused by air flow (\cref{fig:sketches}a).
These predictions of a \emph{dynamic} friction $\zeta(\omega)$ that depends on the oscillation frequency $\omega$ have been interpreted in terms of a delayed response, also referred to as hydrodynamic memory,
and have only recently been shown to be quantitative also for micron-sized particles \cite{Franosch:N2011, Kheifets:S2014}.
In the domain of microrheology, measurements of $\zeta(\omega)$ are used to infer the mechanical properties of complex fluids
\cite{Mason:PRL1995,Gittes:PRL1997,Squires:ARFM2010,Waigh:RPP2016,Rigato:NP2017}.

From the perspective of individual molecules or atoms, fluids are governed by conservative interactions and obey Newton's equations of motion,
yielding smooth and time-reversible trajectories (\cref{fig:sketches}c).
In particular, a single molecule is not subject to friction in this picture, and the mechanism of the required entropy production is far from obvious.
Macroscopic friction and other transport coefficients have been linked to microscopic chaos and the Lyapunov spectrum of the liquid \cite{Dorfman:Chaos, Cohen:PA1995, Posch:PRA1988},
yet the connection of the latter to the corresponding Green--Kubo integrands, or equivalently, to the dynamic friction $\zeta(\omega)$, is an open issue \cite{Cohen:PA1995}. 
First-principle theories to friction are hampered by the fact that liquids are strongly interacting systems.
An insightful, formal relation between the many-particle Liouville operator and dissipation spectra was derived in the seminal works by Zwanzig, Mori, and others \cite{Zwanzig:1965,Mori:1965},
but the analytic evaluation of the resulting expressions hinges on uncontrolled approximations.
For example, a rigorous short-time expansion of the motion at all orders yields zero friction (see Methods).
Early work on dissipation spectra recognised the importance of exact sum rules \cite{Forster:1968,Ailawadi:1971}, the proposed ad-hoc models, however, violate the sum rules at higher orders.
Theoretical progress was made for hard-sphere fluids, where billiard-like collisions generate an instantaneous, Markovian contribution to friction \cite{Bocquet:JSP1994},
thereby rendering the frictionless regime inaccessible.

To gradually bridge between the atomistic and hydrodynamic regimes, one would ideally like to have a magnifying glass that allows for zooming from the slowest to the fastest processes, thus obtaining an increasingly sharper view of the molecular details (\cref{fig:sketches}a).
Spectral quantities such as $\zeta(\omega)$ can serve this purpose with the frequency $\omega$ as the control knob.
Here, we implemented this idea for three distinct liquids and have traced the friction of molecules over wide frequency windows from fully-developed dissipation all the way down to the non-dissipative regime, revealing sizable variations of $\zeta(\omega)$.
Such deviations of $\zeta(\omega)$ from a constant friction $\zeta_0$ signify non-Markovian motion that is widely cast in the generalised Langevin equation [Methods, \cref{eq:GLE}], parametrised by an associated memory function $\gamma(t)$.
The quantities $\zeta(\omega)$ and $\gamma(t)$ are related to each other by a cosine transform, but the determination of either of them
from data is a formidable challenge, with substantial progress in the past years \cite{Shin:CP2010,Gottwald:JCP2015,Lesnicki:PRL2016,Jung:JCTC2017,Meyer:JCP2017,Kowalik:PRE2019,Tassieri:NJP2012,Nishi:SM2018}.
Reaching the high-frequency regime was precluded so far by practical limitations, e.g., statistical noise and insufficient dynamic windows, which we have overcome here by high-precision simulations and an advanced data analysis that utilises physical principles.

For liquid water, for a dense Lennard-Jones (LJ) fluid representing a simple, mono-atomic liquid, and for a supercooled binary mixture serving as a model glass former, we obtained low-noise, cross-validated data for the dynamic friction $\zeta(\omega)$ and the frequency-dependent viscosity $\hat\eta(\omega)$, covering up to 3 orders in magnitude and 4 decades in frequency. Corroborated by these data, we give rigorous arguments that dissipative processes in molecular liquids are exponentially fast
suppressed as frequency is increased. As a consequence, the liquid's response is purely elastic beyond a characteristic frequency $\omega_c$, a feature that goes beyond popular models of friction and viscoelasticity.
The rapid decay of dissipation spectra such as $\zeta(\omega)$ is easily shadowed by approximations,
and as a modelling constraint it is under-investigated in the existing microscopic theories of liquids.
We show further that the high-frequency properties are reproduced by the motion of a single particle in an immobile cage, and our work brings up new, concrete questions on the relation between microscopic chaos and transport
coefficients.
Finally, having data available for both $\zeta(\omega)$ and $\hat\eta(\omega)$, we test the connection between microscopic friction and the macroscopic mechanical properties of complex fluids, postulated by the generalised Stokes--Einstein relation. The latter is found to either fail completely (monoatomic liquid), serve as a qualitative description (water), or being a nearly quantitative relation (supercooled liquid).

\section*{Results} 

\begin{figure}[b]
  \includegraphics[width=\linewidth]{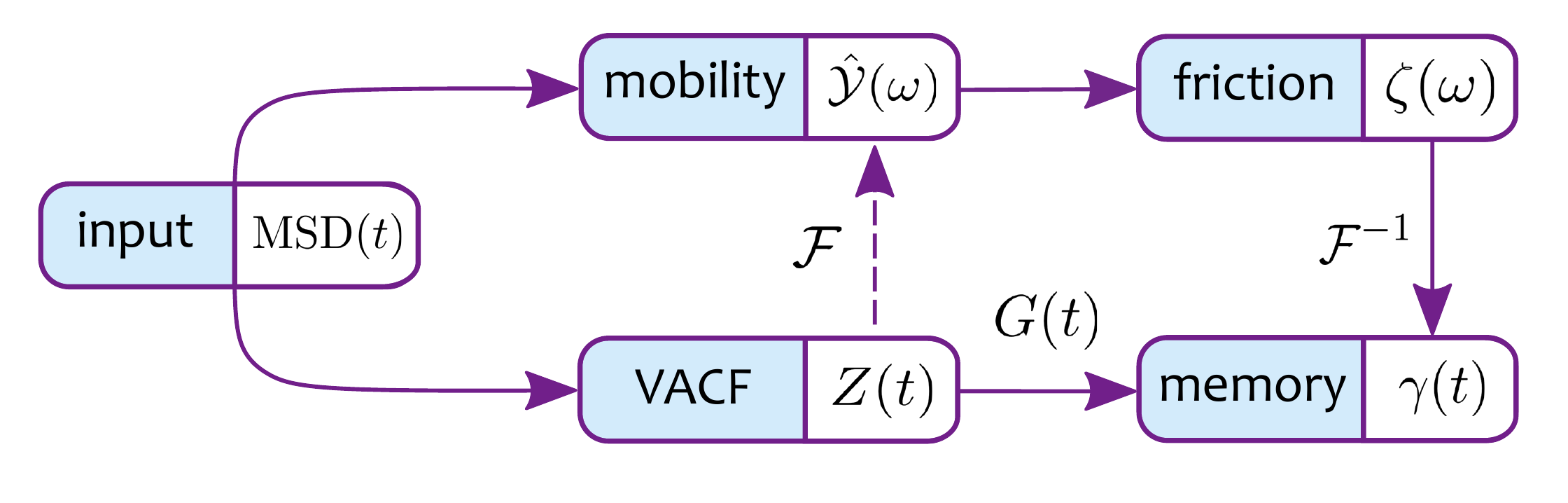}
  \caption[]{
    \textbf{Flow chart of the data analysis.}
    Along the upper route, one starts from the mean-square displacement (MSD) and computes the generalised mobility $\Y(\omega)$
    by numerical differentiation and a suitable Fourier transform (adapted Filon algorithm);
    the dynamic friction $\zeta(\omega)$ follows via \cref{eq:def-friction}.
    A Fourier backtransform then yields the memory function $\gamma(t)$.
    The latter can be obtained more directly along the lower route,
    which is based on the velocity autocorrelation function (VACF) $Z(t)$ and employs a deconvolution in time domain. 
  }
  \label{fig:flowchart}
\end{figure}

\begin{figure*}
  \includegraphics{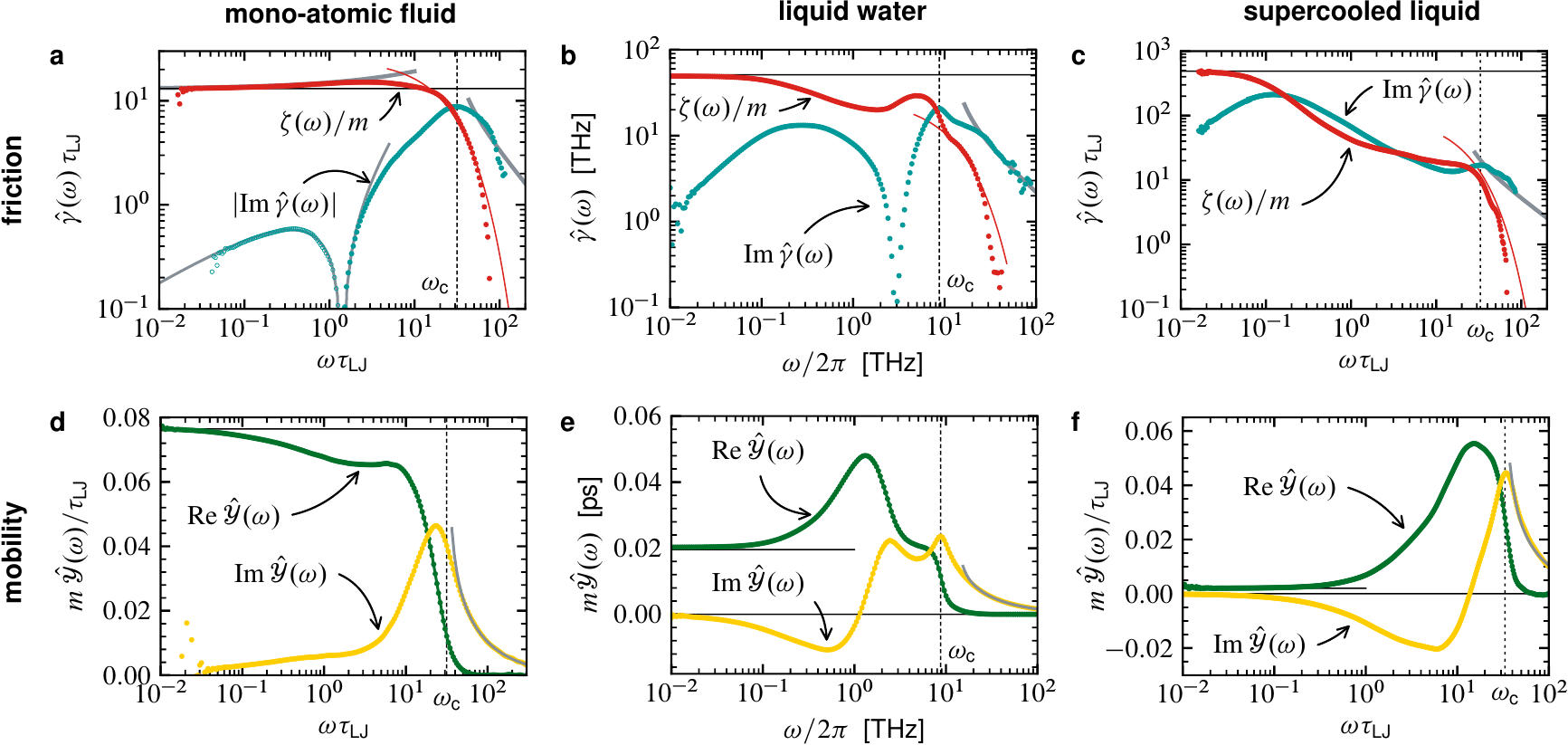}
  \caption[]{
  \textbf{Dynamic friction and generalised mobility of three prototypical liquids}.
  Columns show simulations results for a mono-atomic (LJ) fluid, liquid water, and a supercooled liquid,
  obtained from the data analysis depicted in \cref{fig:flowchart}, with the MSD (not shown) as initial input.
  \:Panels~(a)--(c): The dynamic friction (red) is the real part of the memory kernel $\hat\gamma(\omega)$, as computed from \cref{eq:def-memory2}. It interpolates between the hydrodynamic value $\zeta_0/m$ (horizontal line) and a rapid decrease to zero at high frequencies ($\omega \gg \omega_c$);
  thin red lines mark exponential decays, $\sim \e^{-\omega/\omega_c}$, to guide the eye.
  In case of the LJ fluid (panel~a), $\zeta(\omega)$ is consistent with Stokes's small-$\omega$ asymptote [grey line, \cref{eq:zeta-small-freq}], with parameters taken from a fit to the long-time tail of $Z(t)$ (inset of \cref{fig:results-time}a).
  The elastic response $\Imag \hat\gamma(\omega)$ (turquoise) exhibits a local maximum at high frequencies, defining the characteristic frequency $\omega_c$ (vertical lines), and follows the high-frequency asymptote [\cref{eq:memory_high_freq}, grey line], with all parameters fixed by short-time fits to the MSD data.
  The frequency $\omega_c$ is close to, but different from the Einstein frequency~$\omega_0$.
  \:Panels~(d)--(f): Numerical results for the generalised mobility $\hat{\mathcal{Y}}(\omega)$, which describes the response to an applied small, oscillatory force (see Methods).
  The real part $\Real \hat{\mathcal{Y}}(\omega)$ (green) approaches the hydrodynamic mobility $1/\zeta_0$ (horizontal line) as $\omega \to 0$ and vanishes rapidly at large frequencies.
  The imaginary part (yellow) encodes the elastic response, which has a resonance near $\omega_c$ (vertical line);
  at larger frequencies, the data match with theoretical predictions [grey lines, \cref{eq:mobility_high_freq}].
  }
  \label{fig:results-freq}
\end{figure*}

Instead of observing the response to an oscillatory force, we recorded the Brownian position fluctuations in equilibrium and link them to the friction, taking advantage of a fluctuation dissipation relation.
For the three liquids under investigation, we carried out molecular dynamics simulations to compute
the mean-square displacement (MSD).
Using the MSD as sole input, we estimated both the dynamic friction $\zeta(\omega)$ and the memory function $\gamma(t)$ in an ansatz-free approach, following two complementary routes that allow for cross-validation (see \cref{fig:flowchart} and Methods).
The first route invokes complex analysis and is based on the Fourier--Laplace transform of correlation functions, sampled on a sparse time grid (``adapted Filon algorithm'').
Second and independently, we computed the antiderivative of $\gamma(t)$ using a stable deconvolution technique for uniform time grids.

\subparagraph{Molecular friction in liquids emerges rapidly.}  

Although all three liquids display rather different dynamics, leaving distinct fingerprints in their friction spectra, their high-frequency behaviours of $\zeta(\omega)$ share significant similarities (\cref{fig:results-freq}a--c).
Most remarkably, the data demonstrate that beyond a liquid-characteristic frequency, $\omega \gtrsim \omega_c$, the friction $\zeta(\omega)$ goes exponentially fast to zero. Such a rapid spectral variation has to be contrasted to the typical algebraic peaks, i.e., the Lorentz--Debye shape, and we argue in the following that our finding is generic for molecular fluids.
Upon decreasing frequency, the onset of friction is followed by liquid-specific behaviour over several decades in time until the hydrodynamic value $\zeta_0$ is established:
in water, our results for the friction $\zeta(\omega)$ exhibit a local maximum at $\omega/2\pi \approx \SI{5}{THz}$,
followed by a slow increase towards the limiting value $\zeta_0$, which is reached near frequencies of $\SI{0.1}{THz}$.
For the LJ fluid, $\zeta(\omega)$ varies more smoothly with a global maximum at an intermediate frequency,
and $\zeta_0$ is approached slowly from above in accord with the hydrodynamic square-root singularity [\cref{eq:zeta-small-freq}], essentially calculated by Stokes already \cite{Stokes:1851}.
On theoretical grounds, this feature of $\zeta(\omega)$ is generic for all liquids, yet it is suppressed in our data for the other two liquids due to a small prefactor.
In the supercooled liquid, we observe a scale separation by 3 dynamic decades of \emph{(i)}~the rapid onset of friction and \emph{
(ii)}~the slow emergence of the hydrodynamic limit. The second process is associated with cage relaxation, strongly delayed in the glassy state,
which suggests that the driving mechanism behind the \emph{onset} of friction is not in the structural relaxation of the fluid.
Close to the glass transition, the small-frequency friction $\zeta_0$ is governed by self-similar relaxation processes and obeys asymptotic scaling laws \cite{Fuchs:1998,Goetze:MCT}; the magnitude (prefactor) of these laws, however, is set at high frequencies.

\begin{figure*}
  \includegraphics{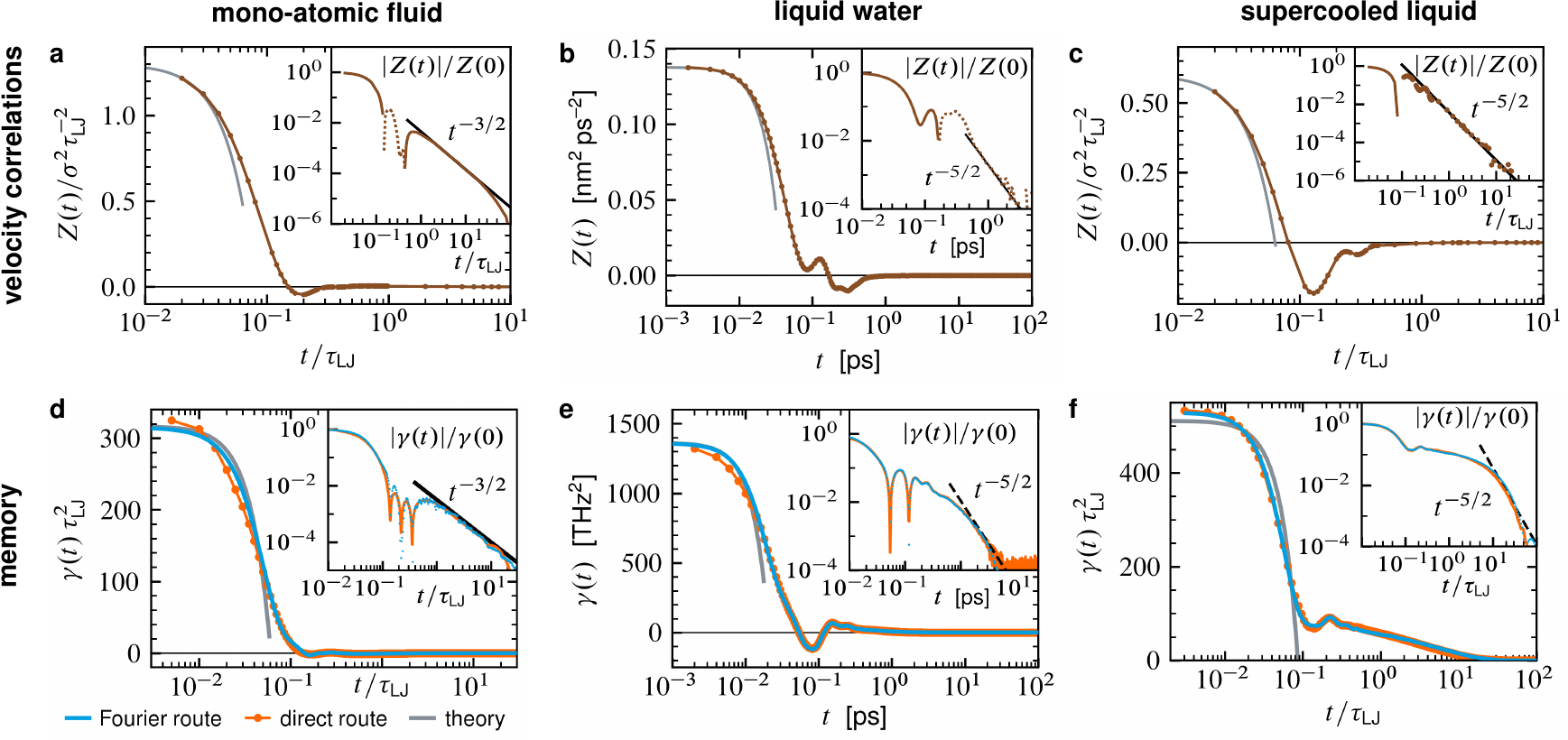}
  \caption[]{
  \textbf{Velocity autocorrelators and corresponding memory functions for the three investigated liquids}.
  Columns show results for a mono-atomic (LJ) fluid, liquid water, and a supercooled liquid.
  \:Panels~(a)--(c): simulation results for the VACF (brown symbols) display a parabolic decrease at short times (grey lines) and power law decays at long times (insets, negative values are dotted).
  The VACF is obtained from the second derivative of the MSD with respect to lag time.
  \:\mbox{Panels (d)--(f):} The memory function $\gamma(t)$ encodes the autocorrelation of Brownian forces on the molecules [\cref{eq:FDT}].
  For each liquid, $\gamma(t)$ was computed from a cosine transform of $\hat\zeta(\omega)$ [blue line, \cref{eq:backtransform}]
  with input data from \cref{fig:results-freq}a--c] and is compared to the deconvolution results in time domain [orange line, \cref{eq:Gi-VACF}].
  The data follow the predicted short-time decay \cref{eq:memory_short_time} (grey line) and exhibit power-law decays
  consistent with \cref{eq:gamma-tails} (insets), preceded by an ultra-slow decay in case of the supercooled liquid [panel~(f)].
  }
  \label{fig:results-time}
\end{figure*}

\subparagraph{Friction depends on the coupling of fast and slow processes.}

The obtained data of $\zeta(\omega)$ cover the full range of the dynamic response, thereby connecting physics at different scales.
Key features will be rationalised by tracing their origins to the dynamics of the fluid particles at short and long times
(going backwards in \cref{fig:flowchart}).
The relevant properties are prominently visible in the second derivative of the MSD, the velocity autocorrelation function (VACF),
$Z(t) := \partial_t^2 \MSD(t)/6$, after numerical differentiation with respect to the lag time $t$ (\cref{fig:results-time}a--c).
The following should be contrasted to Ornstein's model for Brownian motion, 
employing a single exponential decay of velocity correlations,
$Z(t) \approx \vth^2 \exp(-\zeta_0 t /m)$, which implies a constant (Markovian) friction, $\zeta(\omega) \approx \zeta_0$;
by $\vth$ we denote the thermal velocity, and $m$ is the molecular mass.
As a distinct feature of molecular fluids, obeying Newton's equations, the VACF's true short-time decay is parabola-shaped \cite{Hansen:SimpleLiquids}, $Z(t \to 0) \simeq \vth^2 \bigl(1-\omega_0^2 t^2 /2 \bigr)$, introducing
the Einstein frequency $\omega_0$. 
For dense liquids, the VACF, after a sign change, generically displays a regime of anti-correlations, which reflect the transient caging by neighbouring molecules (\cref{fig:sketches}b).
For water and the supercooled liquid, these anti-correlations relax slowly with an intermediate power-law decay, $Z(t) \sim - t^{-5/2}$ (insets of \cref{fig:results-time}b,c);
such a decay was observed earlier in supercooled liquids \cite{Williams:PRL2006,Peng:PRE2016}
and it is a well-established long-time tail
in colloidal suspensions \cite{Ackerson:1982,Fuchs:JPCM2002,Mandal:PRL2019}
and for diffusion in an arrested, disordered environment \cite{vanBeijeren:1982,Hoefling:PRL2007}.

The famous long-time tail encoding hydrodynamic memory \cite{Alder:1967,Ernst:1970,Hansen:SimpleLiquids},
$Z(t \to \infty) \sim t^{-3/2}$, is clearly developed in our data for the LJ fluid after another sign change (inset of \cref{fig:results-time}a),
and in this situation, Stokes's hydrodynamics describes the slow motion of single molecules (\cref{fig:results-freq}a).
For the other two liquids studied, the tail is not visible in our data due to a small prefactor, which
following mode-coupling arguments decreases as either viscosity or diffusivity increases \cite{Hansen:SimpleLiquids}.

\begin{figure*}
  \includegraphics{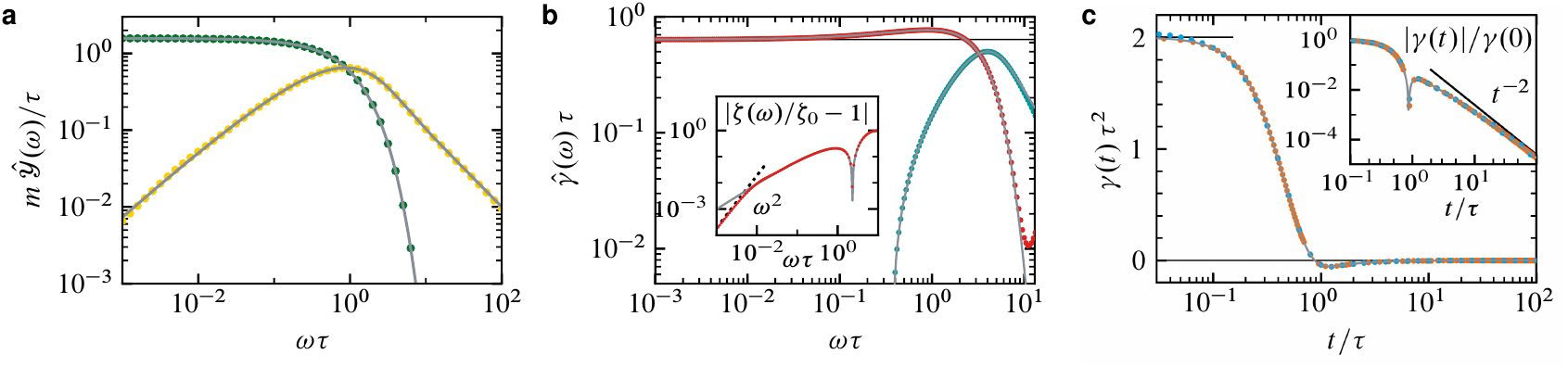}
  \caption[]{
  \textbf{Theoretical model justifies 
  exponentially fast onset of friction and persistent memory.}
  Exact results (grey solid lines) for the VACF model of \cref{eq:VACF-synth}, combining the smoothness and time-reversal symmetry of molecular autocorrelation functions with a long-time tail.
  As a sensitive test of our numerical procedures, symbols show numerical results from the MSD as input, sparsely sampled on a geometrically spaced grid.
  The panels show (a)~the complex-valued, generalised mobility $\Y(\omega)$,
  (b)~the dynamic friction $\zeta(\omega)$ and its elastic counterpart $\Imag \hat\gamma(\omega)$, and
  (c)~the memory function $\gamma(t)$ in time domain.
  The latter inherits the long-time tail $\sim t^{-2}$ from the VACF, but of opposite sign (inset).
  The tail induces non-analytic behaviour of $\zeta(\omega)$ at small frequencies, which crosses over to $\sim \omega^2$ due to the smooth, exponential cutoff of the Fourier integrals (inset of panel b). Same colour code as in \cref{fig:results-freq,fig:results-time}.
  }
  \label{fig:synthetic}
\end{figure*}

The dynamic friction is closely linked to the complex-valued, generalised mobility $\Y(\omega)$
via $\zeta(\omega) = \Real[\Y(\omega)^{-1}]$.
This mobility encodes the response to a small, oscillatory force
and is accessible in, e.g., scattering experiments \cite{Arbe:PRL2016}. 
Here, we computed $\Y(\omega)$ from the VACF upon employing a fluctuation dissipation relation (see Methods and \cref{fig:results-freq}d--f).
Our data reveal a generic, rapid decrease of the dissipative part, $\Y'(\omega) := \Real \Y(\omega)$, upon increasing frequency towards the microscopic regime, $\omega \gg \omega_c$; concomitantly, the elastic response, $\Y''(\omega) := \Imag \Y(\omega)$, has a resonance near $\omega_c$ due to vibrational motion of molecules in their cages.
In the low-frequency limit, the reciprocal of the macroscopic friction is recovered, $\Y(\omega \to 0) = 1/\zeta_0$.
In the examples studied, both regimes are separated by at least two decades in time, which show material-specific features: the mobility of water molecules is 2.5-fold enhanced over its macroscopic value near $\omega/2\pi \approx \SI{1.3}{THz}$; similarly, a factor of 30 is observed for the supercooled liquid.
At variance, the hydrodynamic long-time tail, as for the LJ liquid, demands $\Y'(\omega\to 0)$ to be approached from below.
The interplay of slow and fast processes enters the response $\Y(\omega)$, and thus the dynamic friction $\zeta(\omega)$, at all frequencies, which is borne out by the Kramers--Kronig relations.
In particular, singular behaviour of $\Y''(\omega)$ as $\omega \to 0$ influences the detailed onset of friction at large frequencies.

\subparagraph{Friction is non-local in time.}

The rapid decrease of $\Y'(\omega)$ is mathematically justified from the short-time properties of the VACF.
Physical molecular trajectories, being solutions to Newton's equations, are smooth and yield a smooth function $Z(t)$; in particular, all derivatives $Z^{(n)}(t)$ exist at $t=0$ and are finite.
Thus, invoking exact sum rules [\cref{eq:sum-rules}], \emph{all} moments of the spectrum $\Y'(\omega)$ are finite, which requires an exponentially fast decay as $\omega\to\infty$.
(This is a special case of a more general characterisation of exponentially decaying probability measures\cite{Mimica:JMAA2016}.)
Combining with the large-$\omega$ asymptote of the imaginary part,
$\Y''(\omega) \simeq 1/m\omega \gg \Y'(\omega)$ (see Methods) and using
$
  \zeta(\omega) = \Y'(\omega) / \bigl[\Y'(\omega)^2 + \Y''(\omega)^2 \bigr]
$
proves that $\zeta(\omega) \simeq (m\omega)^2 \, \Y'(\omega)$ as $\omega \to \infty$ and thus an exponentially fast suppression of the friction.
We stress further that such behaviour is not contained in representations of $\zeta(\omega)$ as a truncated continued fraction \cite{Hansen:SimpleLiquids,BoonYip:Molecular}.

It is tempting to use a systematic short-time expansion of $Z(t)$ to predict the large-frequency behaviour of the friction.
However, $Z(t)$ being an even function due to time-reversal symmetry in equilibrium renders the large-$\omega$ asymptotes zero, $\Y'(\omega) \equiv 0$ and thus $\zeta(\omega) \equiv 0$, even if the complete Taylor series of $Z(t)$ in $t=0$ was known
(see Methods).
Note that $\Y''(\omega)$ and the elastic counterpart of $\zeta(\omega)$ are well captured by such an expansion (\cref{fig:results-freq}\mbox{d--i}).
This observation underlines that, on all scales, friction emerges as a phenomenon that is non-local in time, i.e., it cannot be anticipated from the local behaviour of the molecular trajectories.

Our numerical and theoretical findings are supported by an analytically solvable example. The choice
$Z(t) = \vth^2 \bigl[1+(t/\tau)^2\bigr]^{-1}$
combines the smoothness and time reversal symmetry of molecular autocorrelation functions with a long-time tail.
It yields an exponential decay of the mobility, $\Y'(\omega) = (\pi\tau/2m) \,\e^{-|\omega \tau|}$ (dissipative part), and hence a rapid suppression of friction, $\zeta(\omega \to\infty) \sim (\omega \tau)^2 \e^{-\omega\tau}$, as demanded above (see Methods and \cref{fig:synthetic}).

\subparagraph{Irreversible momentum transfer drives the onset of friction.}

\begin{figure}
  \includegraphics[width=\linewidth]{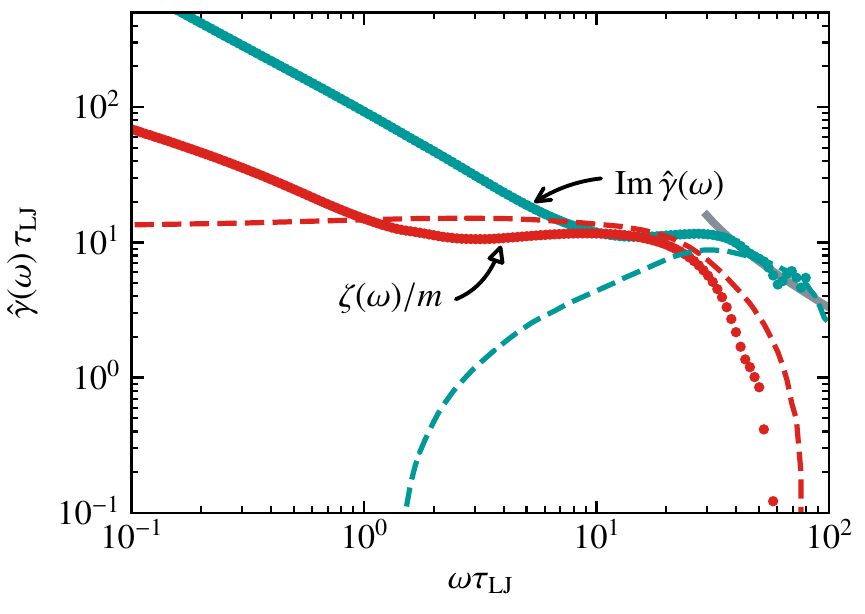}%
  \caption[]{
  \textbf{Friction emerges due to rattling motion in immobile cages.}
  Dynamic friction $\zeta(\omega)$ (red symbols) obtained in a control simulation of a single particle moving in a frozen-in cage formed by neighbouring particles (\cref{fig:sketches}b). The setup was created by pinning the particles of the LJ fluid except for one; results correspond to an ensemble average over \SI{e6} typical cages.
  The imaginary part of the memory kernel $\hat\gamma(\omega)$ (turquoise) ties in with the high-frequency prediction (gray).
  Dashed lines show the results for the unconstrained fluid at the same conditions for comparison (\cref{fig:results-freq}a).
  }
  \label{fig:frozen-cages}
\end{figure}

A pressing question is about the physical mechanism that generates the onset of friction. Motivated by our results for the supercooled liquid (\cref{fig:results-freq}c), we performed a control simulation for the LJ fluid with the structural relaxation switched off by pinning all particles but one.
The rattling motion in such a frozen-in cage (\cref{fig:flowchart})
experiences a dynamic friction that closely resembles our generic findings for $\zeta(\omega)$ at high frequencies, $\omega \gtrsim \omega_c$ (\cref{fig:frozen-cages}).
Upon decreasing frequency from $\omega_c$ to zero, the two dynamics deviate strongly as is most evident in the elastic response:
whereas $\Imag \hat\gamma(\omega)$ decreases for the unconstrained fluid and exhibits a zero crossing enforced by hydrodynamics [\cref{eq:zeta-small-freq-complex} and \cref{fig:results-freq}a], it remains positive for the pinned case and grows as $\Imag \hat\gamma(\omega \to 0) \sim 1/\omega$, reminiscent of what one obtains for an Ornstein--Uhlenbeck (OU) particle in a harmonic trap \cite{Franosch:N2011}.
For even smaller frequencies, $\omega \lesssim \tau_\text{LJ}^{-1}$, the dissipation diverges too, approximately as $\zeta(\omega \to 0) \sim \omega^{1/2}$, which we attribute to the irregular shape of the confining cages; for the OU model with harmonic confinement, $\zeta(\omega) \simeq \mathit{const}$.
At high frequencies, however, the confinement is not relevant, and 
we conclude that it is the fast, yet irreversible momentum transfer to neighbouring molecules that drives the onset of friction.
This is corroborated by the observation that instantaneous momentum exchange implies a non-zero limit, $\zeta(\omega \to \infty) > 0$, as in the case of hard spheres \cite{Bocquet:JSP1994}.
The fact that dissipation is linked to trajectories for which the time-reversed path is extremely improbable
leads us to speculate that the onset frequency $\omega_c$ is intimately related to the largest Lyapunov exponent $\lambda_\mathrm{max}$ of the fluid, which is close to, but different from the Einstein frequency $\omega_0$ (ref.\citenum{Posch:PRA1988}).

\subparagraph{Dynamic friction implies intricate memory of Brownian motion.}

Within the framework of the generalised Langevin equation [\cref{eq:GLE}], momentum relaxation is governed by a memory function $\gamma(t)$ that is fully determined by $\zeta(\omega)$ [\cref{eq:def-friction,eq:backtransform}].
At the same time, $\gamma(t)$ is also the autocorrelator of Brownian, random forces on the molecules, up to a constant prefactor, and $\zeta(\omega)$ encodes the corresponding spectrum [\cref{eq:FDT}].
Within Ornstein's idealised model of Brownian motion, one assumes independent Brownian forces, implying
a flat, white spectrum, $\zeta(\omega) \approx \zeta_0$, and a delta-peaked memory function $\gamma(t)$.
For molecular liquids, however, the memory functions display a universal parabola-shaped short-time decay [\cref{fig:results-time}d--f and \cref{eq:memory_short_time}].
For water and the LJ fluid, the short-time regime of $\gamma(t)$ is followed by oscillatory behaviour including sign changes and, finally, different power-law decays for the two liquids encoding different physics (insets of \cref{fig:results-time}d,e, also see \cref{fig:synthetic}c);
generally, power-law tails of the memory function are directly inherited from the VACF without a change of exponent [\cref{eq:gamma-tails}].
For the supercooled liquid, $\gamma(t)$ remains positive and exhibits the onset of a plateau (near $t \approx 0.3\tau_\mathrm{LJ}$), which decays logarithmically slowly over 2 decades in time (\cref{fig:results-time}f).
From the modelling perspective, it is desirable to approximate the memory functions such that the initial decay, the typical persistence time, and the integral of $\gamma(t)$ are reproduced, the latter yielding $\zeta_0$ [\cref{eq:friction_Green_Kubo}].
For all three liquids, the complexity and the long-lived nature of the memory, however, preclude simple models of $\gamma(t)$ such as the superposition of few exponential decays.
The quantitative knowledge of $\zeta(\omega)$, as obtained here, paves the way for more favourable approximations of memory in the frequency domain, which can yield mathematically and physically consistent interpolations of Brownian motion from the fastest to the slowest scales.

\section*{Discussion}

\subparagraph{Relation to viscosity and elasticity of liquids.}

Going beyond the dynamics of single molecules and their friction, the relation to the visco-elastic properties of complex fluids
is of ongoing interest for the physics of polymers, living cells, and the glass transition.
The potentially tight coupling between single-particle and collective responses was phrased as an \emph{ad hoc} extension of Stokes's friction law to the frequency domain,
$
  \zeta(\omega) = 6\pi \Real[\hat\eta(\omega)] a,
$
referred to as generalised Stokes--Einstein relation (GSER),
which has found wide applications in the context of microrheology experiments
\cite{Mason:PRL1995,Gittes:PRL1997,Squires:ARFM2010,Waigh:RPP2016}.
It links the dynamic friction $\zeta(\omega)$ of a probe particle to the dynamic shear modulus,
$\what G(\omega) = -\i\omega \hat\eta(\omega)$,
a complex-valued function encoding the stress response of the macroscopic fluid sample to a small, oscillatory shear strain.
The generalised viscosity $\what \eta(\omega)$ tends to the hydrodynamic shear viscosity $\eta_0$ as $\omega \to 0$,
with $\Real \what \eta(\omega)$ representing the spectrum of shear stress fluctuations (up to a constant factor) by a fluctuation dissipation relation.
Thus, if the GSER holds the single-particle memory $\gamma(t)$ is proportional to the autocorrelator of shear stresses, which means that the Brownian force on the particle and the fluctuation of the shear stress are statistically equivalent variables.

\begin{figure*}
  \includegraphics[width=\linewidth]{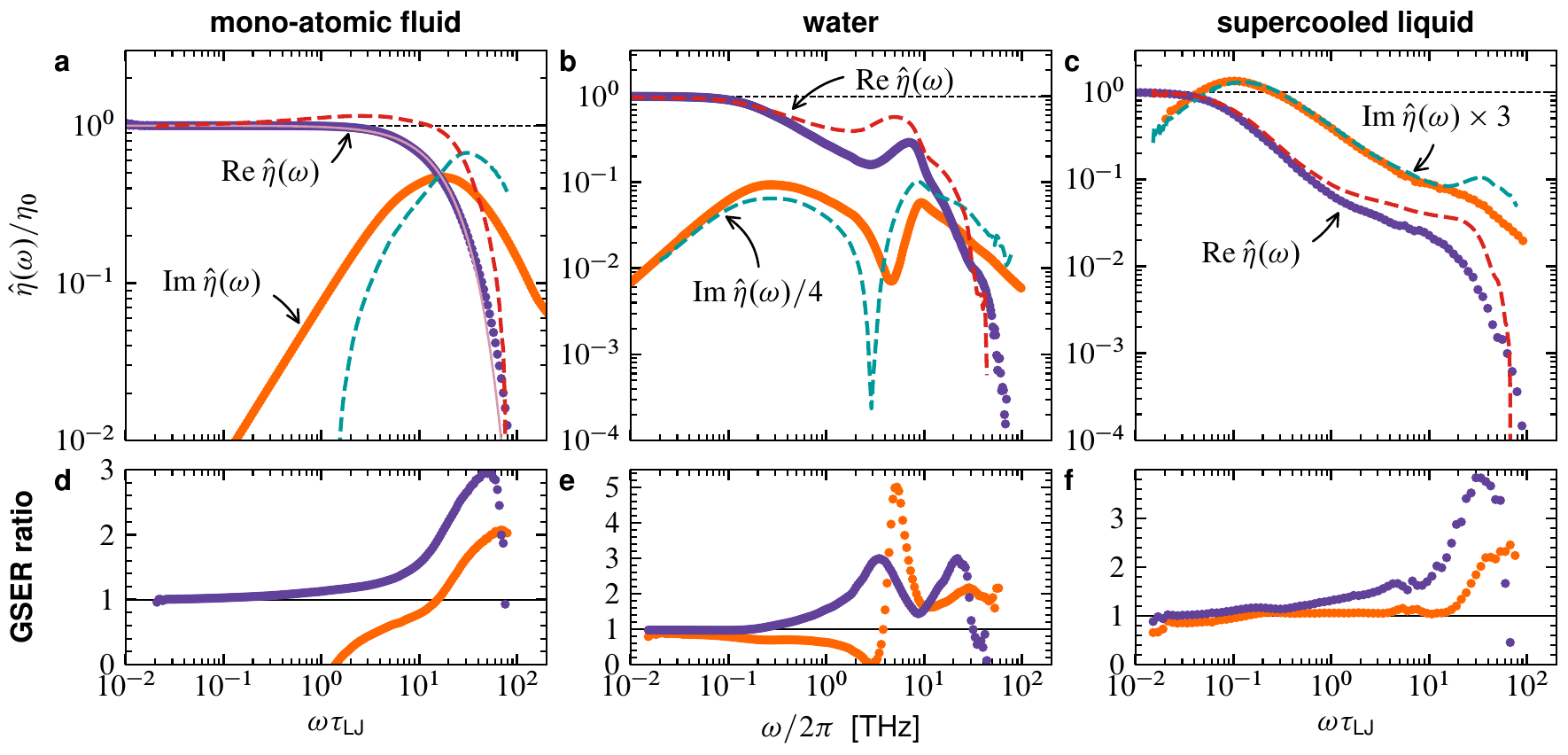}
  \caption{
  \textbf{Test of the generalised Stokes--Einstein relation (GSER).}
    Panels~(a--c): The generalised viscosity $\Real \hat\eta(\omega)$ (violet symbols) of the three liquids under investigation (columns) is compared to the GSER prediction $\zeta(\omega) / 6\pi a$ (red dashed lines), based on the dynamic friction data of single molecules (\cref{fig:results-freq}a,c), and correspondingly for the imaginary counterparts of the elastic responses (orange symbols and turquoise dashed lines).
    The effective particle radius $a$ for each liquid is chosen such that the viscosity and friction curves coincide at $\omega = 0$.
    The pink solid line in panel~(a) is an empirical fit of a compressed exponential, $\Real \hat\eta(\omega) \simeq \eta_0 \exp(-(\omega/\omega_\eta)^\beta)$ with $\beta=1.29$ and $\omega_\eta = 1.19\,\omega_0$.
    In panels~(b,c), the data for the elastic responses are shifted by the indicated factors for clarity.
    Panels~(d--f): The GSER is tested by plotting the ratios $\zeta(\omega)/6\pi [\Real \hat\eta(\omega)] a$ (violet) and
    $m [\Imag \hat\gamma(\omega)]/6\pi [\Imag \hat\eta(\omega)] a$ (orange); deviations from unity quantify the GSER violation.
  }
  \label{fig:viscosity}
\end{figure*}

A critical assessment of the validity of the GSER is permitted by comparing our data for $\zeta(\omega)$ to results for $\hat\eta(\omega)$, calculated within the same simulations (see \cref{fig:viscosity} and Methods).
Generically, the dissipative part, $\Real \hat\eta(\omega)$, decays exponentially fast as $\omega \to \infty$, which is required by analogous arguments as given for $\zeta(\omega)$ and $\Y'(\omega)$.
For high frequencies, only the imaginary part survives due to its slow decay, $\hat\eta(\omega) \simeq G_\infty/(-\i\omega)$, inducing a non-zero and real-valued modulus, $\hat G(\omega \gg \omega_c) \approx G_\infty > 0$.
Therefore, our data clearly demonstrate that, indeed, liquids respond to high-frequency shear like a non-dissipative, elastic solid as put forward by the classical work of Frenkel \cite{Frenkel:KineticTheory}.
However, the attempt to predict the elastic modulus $G_\infty$ from the vibrational motion of molecules in their cages, by virtue of the GSER, would considerably overestimate the modulus by factors of $\approx 2$ for the three liquids studied (\cref{fig:viscosity}e--f).
In passing, we note that Maxwell's model for viscoelasticity \cite{Frenkel:KineticTheory,BoonYip:Molecular,Hansen:SimpleLiquids},
$\hat\eta(\omega) = G_\infty\tau/(1 - \i\omega\tau_M)$ with some relaxation time $\tau_M$, breaks down at high frequencies as it implies a slowly decaying real part, $\Real \hat\eta(\omega \to\infty) \sim \omega^{-2}$, in sharp contrast to exact sum rules \cite{Forster:1968} and to the exponentially fast decay for molecular liquids. Therefore, treatments of sound-like, elastic waves on the footing of this and similar models appear incomplete.

The passage from the elastic to the viscous limit occurs upon decreasing frequency, leading in case of the LJ fluid to a monotonic increase of $\Real \hat\eta(\omega)$, which, empirically, follows a compressed exponential over almost the full frequency domain (\cref{fig:viscosity}a).
In particular, $\hat\eta(\omega) \approx \eta_0$ is constant for $\omega \lesssim 2\tau_\mathrm{LJ}^{-1}$, which defines the hydrodynamic regime.
For these frequencies, the single-molecule response is very well described by Stokes's dynamic friction [\cref{eq:zeta-small-freq}],
making the GSER violation apparent for Newtonian fluids, for which $\hat\eta(\omega) = \eta_0$.
It is evidenced further by the dissimilarity of the elastic parts, $\Imag \hat\gamma(\omega)$ and $\Imag \hat\eta(\omega)$.


For water, the viscosity and friction spectra share similar features and coincide fairly well (\cref{fig:viscosity}b), including the elastic parts. Thus, the GSER serves as a reasonable qualitative description, in particular for frequencies below $\approx \SI{2}{THz}$, i.e., slower than the vibrations of the first hydration shell. A detailed analysis of the visco-elastic spectrum of water can be found in ref.~\citenum{Schulz:2020}.

In supercooled liquids, the Stokes--Einstein relation for molecules (i.e., the GSER for $\omega\to 0$) holds in the presence of a huge variation of the viscosity. In particular, the ratio $\zeta_0 / \eta_0$ is observed to be constant over a wide temperature range---in line with the mode-coupling theory of the idealised glass transition \cite{Goetze:MCT}.
At very low temperatures, however, marked deviations from the Stokes--Einstein relation (mostly studied for $\omega \to 0$) have received considerable attention as they signify the opening of additional relaxation channels not included in the standard theory \cite{Tarjus:JCP1995,Kumar:JCP2006,Gupta:PRL2015,Dehaoui:PNAS2015,Parmar:PRL2017}.
For the moderately supercooled liquid studied here exemplarily, both viscous and elastic responses satisfy the GSER over a wide frequency window (\cref{fig:viscosity}c).
Notably, the elastic components collapse almost perfectly in this case, $\Imag \hat\gamma(\omega) \sim  \Imag \hat\eta(\omega)$, which we attribute to the same (apparent) power law scaling, $\approx \omega^{-0.75}$, at intermediate frequencies.
Yet, the collective response lacks the elastic peak of $\Imag \hat\gamma(\omega)$ at $\omega_c$,
causing the breakdown of the GSER at large frequencies.
This suggests that a future frequency-resolved study of systematic deviations from the GSER upon further supercooling can clarify the separate contributions of fast and slow processes to the decoupling of diffusion and viscosity (``Stokes--Einstein violation'') close to the glass transition temperature.

\subparagraph{Further perspectives.}

Molecular friction in liquids arises from a complex interplay of processes on disparate time scales,
and the large variability of $\zeta(\omega)$ over orders of magnitude reveals the strongly non-Markovian nature of Brownian motion
in liquid environments, with far-reaching implications for nanoscale processes.
Examples are as diverse as
reaction rates and barrier crossings in macromolecular dynamics \cite{Guerin:N2016,Sancho:NC2014,Daldrop:PNAS2018}
and flows near liquid--solid interfaces \cite{Mo:JCP2017,Huang:NC2015,Bocquet:SM2007,Fuchs:JPCM2002};
the ability to quantify the corresponding memory is vital for their realistic descriptions.

Beyond that, the finding of a generic drop of $\zeta(\omega)$ at a large, liquid-specific frequency $\omega_c$ marks the rapid onset of friction, which we attribute to the momentum transfer to neighbouring molecules.
These results refine the fundamental question on a quantitative link between friction and microscopic chaos: Whether and how does the frequency-dependence of transport coefficients relate to the Lyapunov spectrum of the liquid?\cite{Cohen:PA1995,Dorfman:Chaos}

From a numerical point of view, our ansatz-free approach
has immediate applications to and extends current methods \cite{Nishi:SM2018,Tassieri:NJP2012} for
the analysis of high-resolution microrheology data \cite{Franosch:N2011,Kheifets:S2014,Berner:NC2018,Waigh:RPP2016},
which involves deducing frequency-dependent moduli from the displacement statistics
along the same lines as done here for the dynamic friction.
Relying merely on the existence of a steady state [cf.\ \cref{eq:memory-general}], the developed methodology is not limited to friction, but can be transferred to the analysis of non-Markovian time series from simulation and experiment.
It finds novel uses in, e.g., the anomalous diffusion within living cells \cite{Hoefling:RPP2013} and the kinetics of chemical reactions \cite{Herrera-Delgado:PCB2018}. It opens a promising avenue for research on the migration of malignant cells in tissue \cite{Hakim:RPP2017} and on predictive stochastic models of financial market \cite{Kanazawa:PRL2018} and geographic  \cite{Franzke:WCC2014} data.

\def\bibname{References}
\bibliography{dynamic-friction}

\clearpage
\let\clearpage\relax 

\section*{Methods} 


\subparagraph{Generalised Langevin equation.}

A labelled fluid particle of mass $m$, position $\vec r(t)$, and momentum $\vec p(t) = m\dot{\vec r}(t)$ obeys the generalised Langevin equation (GLE) \citeM{Kubo:Statistical_Physics_II}: 
\begin{equation}
  \dot{\vec p}(t) = - \int_0^t \! \text{d}s \, \gamma(t-s) \, \vec p(s) + \vec\xi(t),
  \label{eq:GLE}
\end{equation}
where the Brownian force $\vec\xi(t)$ is a stochastic process with zero mean and covariance
\begin{equation}
  \langle \vec\xi (t) \otimes \vec \xi(t')\rangle = m\kB T \gamma(|t-t'|)\, \mathbf{1} \,
  \label{eq:FDT}
\end{equation}
to satisfy the fluctuation--dissipation theorem.
Rewriting \cref{eq:GLE} for the
the velocity auto-correlation function (VACF),
$Z(t) = \langle \vec p(t) \cdot \vec p(0) \rangle / (m^2 d)$, describing momentum relaxation,
yields
\begin{align}
\dot Z(t) = - \int_0^t \text{d}s \, \gamma(t-s) Z(s)\,, \quad Z(0)=\frac{k_\text{B}T}{m}\,.
\label{eq:VACF}
\end{align}
Its Fourier--Laplace transform [\cref{eq:FLtrafo}] provides the link to and the definition of the (complex-valued) memory kernel $\hat\gamma(\omega)$,
\begin{equation}
  \what Z(\omega) = \frac{\kB T/m}{-\i \omega + {\hat\gamma(\omega)}}.
  \label{eq:def-memory}
\end{equation}

\clearpage
\subparagraph{Linear response.}
For a mass $m$ driven by a periodic force $\vec F(t) = \vec F_\omega \, \cos(\omega t)$ with frequency $\omega$ and amplitude $\vec F_\omega$,
the stationary response $\bar{\vec v}(t)$, averaged over many realisations of the experiment, obeys \citeM{Kubo:Statistical_Physics_II}
\begin{equation}
  m \frac{\diff}{\diff t} \bar{\vec v}(t)
    = \vec F(t) - \int_{-\infty}^t \! m \gamma(t-s) \, \bar{\vec v}(s) \, \diff s\,,
  \label{eq:GLE-response}
\end{equation}
corresponding to \cref{eq:GLE}
after shifting the lower integral boundary to $-\infty$ to ensure relaxation of transients.
The upper boundary can be shifted to $+\infty$ with the convention that the response function $\gamma(t < 0) = 0$.
By linearity of the equation, the solution is of the form $\bar{\vec v}(t) = \Real \bigl[ \vec v_\omega \e^{-\i\omega t} \bigr]$
with complex amplitude $\vec v_\omega$, and inserting into \cref{eq:GLE-response} yields
$\vec v_\omega = \Y(\omega) \, \vec F_\omega$
in terms of the generalised mobility,
\begin{equation}
  \Y(\omega) := [-\i\omega m + m\hat \gamma(\omega)]^{-1} \,,
  \label{eq:def-memory2}
\end{equation}
also referred to as complex-valued admittance.
Its central ingredient is the one-sided Fourier transform of the response function,
$\what\gamma(\omega) := \int_0^\infty \e^{\i\omega t} \gamma(t) \,\diff t$.
Comparing to \cref{eq:def-memory}, which describes equilibrium correlations, yields the fluctuation--dissipation relation: $\what Z(\omega) = \kB T \, \Y(\omega)$.

Friction describes the resistance to a prescribed velocity, as in Stokes's pendulum experiments \cite{Stokes:1851}.
Thus, inverting the above argument, the force response to an oscillatory velocity $\vec v(t) = \vec v_\omega \cos(\omega t)$ has complex amplitude $\vec F_\omega = \Y(\omega)^{-1} \, \vec v_\omega$,
and we identify $\Y(\omega)^{-1}$ as a generalised friction.
However, merely the real part describes dissipation and deserves to be called a friction,
which is seen from the mean dissipated power:
$T_p^{-1} \int_0^{T_p} \vec v(t) \cdot \vec F(t) \, \diff t = \Real \bigl[\Y(\omega)^{-1}\bigr] \, |\vec v_\omega|^2 / 2$,
averaged over a full cycle of length $T_p = 2\pi/\omega$.
Thus, we set the dynamic friction as
\begin{equation}
  \zeta(\omega) := \Real \bigl[\Y(\omega)^{-1}\bigr] = m \Real \hat\gamma(\omega) \,;
  \label{eq:def-friction}
\end{equation}
in particular, $\zeta(\omega) \geq 0$.
This is in line with the conventional (Markovian) Langevin equation, $\dot{\vec p}(t) = -(\zeta_0/m) \, \vec p(t) + \vec \xi(t)$.
There, the response is governed by $\Y(\omega) = [-\i\omega m + \zeta_0 ]^{-1}$, implying a static friction,
$\zeta(\omega) = \zeta_0$.

\Cref{eq:def-memory2,eq:def-friction} (and variants thereof) are the basis of (passive) microrheology experiments
\cite{Mason:PRL1995,Gittes:PRL1997,Squires:ARFM2010,Waigh:RPP2016},
which use observations of Brownian motion to infer the friction $\zeta(\omega)$ and $\Imag \hat\gamma(\omega)$ of a probe particle in a complex medium and relate it via the GSER to the local visco-elastic properties.
The macroscopic shear viscosity, $\eta_0 = (\kB T)^{-1} \int_0^\infty C_\Pi(t) \,\diff t$, is the Green--Kubo integral of the autocorrelation,
$C_\Pi(t) = \langle \delta \Pi^\perp(t) \, \delta \Pi^\perp(0)\rangle /V$,
of shear stress fluctuations $\delta \Pi^\perp(t)$, given as an off-diagonal element of the stress tensor \cite{Hansen:SimpleLiquids}; $V$ denotes the sample volume.
Similarly by a fluctuation--dissipation relation, the frequency-dependent response coefficient $\hat\eta(\omega)$ to oscillatory shear is the Fourier--Laplace transform [\cref{eq:FLtrafo}] of $C_\Pi(t)/\kB T$, and thus $\hat\eta(\omega \to 0) = \eta_0$.

\clearpage
\subparagraph{Mathematical framework.} 

For the harmonic analysis of the autocorrelation function $C(t)$ of a stationary time series, we use the Fourier--Laplace transform
\begin{equation}
  \what C(z) = \int_0^\infty \! \e^{\i z t} C(t) \, \diff t \,,
  \label{eq:FLtrafo}
\end{equation}
which is well-defined for all frequencies $z$ in the upper complex plane, $\mathbb{C}_+ = \{z|\Imag z > 0\}$.
Along the imaginary axis, $z=\i y$, it recovers the conventional Laplace transform.
For real frequencies $\omega$, the real and imaginary parts of $\what C(\omega)$ describe physically accessible spectra,
which are related to each other by Kramers--Kronig integrals \citeM{Kubo:Statistical_Physics_II};
for example, $\Real \what C(\omega)$ for fixed $\omega$ is determined by the full function $\Imag \what C(\omega)$.
The real part is positive, $\Real \what C(\omega) \geq 0$, and most importantly, we have the unique Fourier backtransform:
\begin{equation}
  C(t) = \frac{1}{\pi}\int_{-\infty}^\infty \! \e^{-\i\omega t} \Real \what C(\omega) \,\diff \omega \,.
  \label{eq:backtransform}
\end{equation}
If $C(t)$ is $n$-times continuously differentiable at $t=0$, this implies sum rules for the spectrum ($k=0, 1, \dots, n$):
\begin{equation}
  \frac{1}{\pi} \int_{-\infty}^\infty (-\i\omega)^k \Real [\what C(\omega)]\,\diff \omega = C^{(k)}(0) < \infty \,.
  \label{eq:sum-rules}
\end{equation}
In equilibrium, only positive frequencies are needed as $\Real \what C(\omega) = \Real \what C(-\omega)$,
and the integrals are real-valued.

Next, we introduce a memory function of $C(t)$ solely by invoking results from complex analysis \citeM{Teschl:MathMeth,Franosch:JPA2014}.
$\what C(z)$ as above is a holomorphic function with $\Real \what C(z) \geq 0$,
i.e., $\i \what C(z)$ is of Herglotz--Nevanlinna type,
and $(\Imag z) \, |\what C(z)|$ bounded in $\mathbb{C}_+$.
Suppose that $C(t)$ has a regular short-time expansion,
$\textstyle C(t \to 0) \simeq C_0 \bigl[1 - \nu t - \frac{1}{2}a t^2\bigr]$, which implies
\begin{equation}
  \what C(z) \simeq C_0 (-\i z)^{-1} - \nu C_0 (-\i z)^{-2} - a C_0 (-\i z)^{-3}
\end{equation}
for large frequencies, $|z| \to \infty$ with $|\!\arg z| > \delta$ for some $\delta > 0$.
Under these mild requirements, one shows \citeM{Franosch:JPA2014}:
For given $\what C(z)$ there is a unique memory kernel $\what M(z)$
such that
\begin{equation}
  \what C(z) = \frac{C_0}{-\i z + \what M(z)}
  \label{eq:memory-general}
\end{equation}
with $\i\what M(z)$ of Herglotz--Nevanlinna type
and $\what M(z) \simeq \nu + a / (-\i z)$ as $|z| \to \infty$.
In particular, $\what M(z)$ corresponds to the autocorrelation function of another (\emph{a priori} unknown) observable.
Iterating the argument yields the continued-fraction representation of $\what C(z)$, well-known from the Zwanzig--Mori projection formalism \cite{Hansen:SimpleLiquids}.

In the context of the VACF, one puts $C_0 = \vth^2$, $\nu = 0$, and $a = \omega_0^2$ and infers for the memory kernel $\what M(z) =: \hat\gamma(z)$ that $\Real \hat\gamma(z) \geq 0$ and $\hat\gamma(z) \simeq \omega_0^2/(-\i z)$ as $|z|\to\infty$.
This justifies \cref{eq:def-memory} independently of the notion of a GLE,
after taking $z$ along the real line.

By means of \cref{eq:backtransform}, $\hat\gamma(z)$ specifies the memory function $\gamma(t)$, which has a physical interpretation as the autocorrelator of the fluctuating acceleration $\vec \xi(t)/m$ in \cref{eq:GLE}, divided by $\vth^2$.
At low frequencies, $m\hat\gamma(z \to 0) = \zeta_0$ implies a Green--Kubo relation for the hydrodynamic friction:
\begin{equation}
  \zeta_0 = m \int_0^\infty \! \gamma(t) \,\diff t \,.
  \label{eq:friction_Green_Kubo}
\end{equation}

\clearpage
\subparagraph{Short-time expansion.}

The smoothness of physical molecular trajectories, being solutions to Newton's equations, allows for a rigorous short-time expansion of the VACF. Combining with the time-reversal symmetry in equilibrium, $Z(t) = Z(-t)$, only even powers in $t$ contribute and one obtains $Z(t \to 0) \simeq \kB T \sum_{k=0}^{\infty} c_k t^{2k} / (2k)!$ with Taylor coefficients $c_k$ given from equilibrium matrix elements of powers of the underlying Liouville operator \cite{BoonYip:Molecular}.
To connect with the notation of the main text, $c_0 = 1/m$, $c_1 /c_0 = - \omega_0^2$, and we put $c_2 /c_0 =: \Omega^4$.
Fourier--Laplace transforming term by term yields the high-frequency expansion of $\what Z(\omega)$ and
thus of $\Y(\omega)=(\kB T)^{-1} \what Z(\omega)$, which is purely imaginary:
$\Y(\omega \to \infty) \simeq \sum_{k=0}^{\infty} c_k \, (-\i\omega)^{-1-2k} = c_0/(-\i\omega) + c_1/(-\i\omega)^3 + \dots$.
Using \cref{eq:def-friction},
we have $\zeta(\omega) = |\Y(\omega)|^{-2} \Real \Y(\omega)$,
which implies that for high frequencies the friction vanishes, $\zeta(\omega) \equiv 0$, at all orders in $\omega\to \infty$.
A similar situation is familiar from calculus text books: $f(x)=\e^{-1/x}$ has a Taylor series $f(x) \equiv 0$ at $x=0$;
the radius of convergence is~0.

The expansion of $\Y(\omega)$ can be represented as a continued fraction that has the same large-$\omega$ asymptotics up to terms of order $\omega^{-5}$:
\begin{equation}
  \Y(\omega) \simeq 1 \big / \bigl\{ -\i\omega m
    + m \omega_0^2 \big / \bigl[-\i\omega + \omega_1^2 / (-\i\omega + \dots) \bigl] \bigr\} \,,
  \label{eq:mobility_high_freq}
\end{equation}
introducing $\omega_1^2 := \bigl(\Omega^4-\omega_0^4\bigr) \bigl/ \omega_0^2$ for brevity.
This truncation is an excellent description of our data for $\Y''(\omega)$ at high frequencies,
with $\omega_0$ and $\Omega$ obtained from fits to $Z(t)$, see \cref{fig:results-freq}d--f.
For the memory kernel $\hat\gamma(\omega)$, one reads off
\begin{equation}
  \hat\gamma(\omega \to \infty) = \omega_0^2/(-\i \omega)
    - \omega_0^2 \omega_1^2 / (-\i\omega)^3 + \mathcal{O}\bigl(\omega^{-5}\bigr) \,,
  \label{eq:memory_high_freq}
\end{equation}
using \cref{eq:def-memory2},
implying for the memory function in time domain:
\begin{equation}
  \gamma(t \to 0) = \omega_0^2 \left[1 - \omega_1^2 t^2/2 \right] + O\bigl(t^4\bigr).
  \label{eq:memory_short_time}
\end{equation}

\clearpage
\subparagraph{Long-time tails.}

In an unbounded fluid, momentum conservation leads to persistent velocity correlations,
$Z(t \to \infty) \simeq \vth^2 (t/\tau_\infty)^{-3/2}$,
which was explained in terms of hydrodynamic backflow and diffusion of a momentum vortex \cite{Alder:1967,Ernst:1970,Hansen:SimpleLiquids}.
The tail induces a small-$\omega$ singularity in the frequency domain \citeM{Karamata:1931}, which reads for the memory kernel:
$m \hat \gamma(\omega \to 0) \simeq \zeta_0 \bigl[1 + (\tau_\infty \zeta_0/m) \sqrt{-4\pi\i\omega \tau_\infty}) \bigr]$,
using \cref{eq:def-memory} and the hydrodynamic friction $\zeta_0 := m\hat\gamma(0) = \kB T \big/ \int_0^\infty Z(s)\,\diff s$.

In the framework of the creeping flow equations, Stokes found \cite{Stokes:1851}
\begin{equation}
  m \hat\gamma(\omega \to 0) \simeq 6\pi\eta_0 a(1+\sqrt{- \i\omega \tau_\mathrm{fl}}) -\i\omega m_\mathrm{fl}/2
  \label{eq:zeta-small-freq-complex}
\end{equation}
in terms of the vorticity diffusion time $\tau_\mathrm{fl}$
and an effective particle mass $m_\mathrm{fl}$;
matching with the previous expression for the asymptote of $\hat \gamma(\omega)$, one identifies
$\tau_\mathrm{fl} = 4\pi (\zeta_0/m)^2 \tau_\infty^{3}$.
The real part yields the dynamic friction,
\begin{equation}
  \zeta(\omega \to 0) \simeq 6\pi\eta_0 a (1+\sqrt{\omega \tau_\mathrm{fl}/2}) \,,
  \label{eq:zeta-small-freq}
\end{equation}
showing that its macroscopic limit, $\zeta_0 = 6\pi\eta_0 a$, is approached from above as $\omega \to 0$ (see \cref{fig:results-freq}b).
For water and the supercooled liquid, a different type of power law decay, $Z(t) \sim - t^{-5/2}$, was found
(\cref{fig:results-freq}e,f).

For a general long-time tail of the VACF, $Z(t) \sim t^{-\sigma}$ with $\sigma > 1$,
the memory function $\gamma(t)$ asymptotically inherits a power-law decay with the same exponent, but of opposite sign \citeM{Corngold:PRA1972}:
\begin{equation}
  \gamma(t) \simeq -\hat\gamma(0)^2 Z(t) / Z(0) \,, \quad t \to \infty\,;
  \label{eq:gamma-tails}
\end{equation}
which follows from \cref{eq:VACF} and by invoking a Tauber theorem \citeM{Karamata:1931}.
Without any adjustable parameter, the prediction is in excellent agreement with our data for $\gamma(t)$ in case of the LJ fluid.
Inspection of a few examples (figs.~\ref{fig:results-time}d,e and \ref{fig:synthetic}c) suggests that, in order to accommodate the sign change of the tail, the number of zero crossings (knots) in $\gamma(t)$ is increased by one relative to $Z(t)$.

\clearpage
\subparagraph{Analytically solvable example.}
Consider the following analytically tractable model for the VACF:
\begin{equation}
  Z(t)= \frac{\vth^2}{1+(t/\tau)^2} \,, \quad \vth^2 = \kB T/m\,,
  \label{eq:VACF-synth}
\end{equation}
with relaxation time $\tau$ and thermal velocity $\vth$ (supplemental fig.~S1d).
It favourably combines the physically required smoothness at $t=0$ and time-reversibility, $Z(t) = Z(-t)$,
with a power-law decay at long times,
$Z(t\to\infty) \simeq \vth^2 (t/\tau)^{-2}$;
in particular, only even powers of $t$ contribute to the the short-time expansion:
$Z(t\to 0) \simeq \vth^2 \left[1 - (t/\tau)^2 + O(t^4) \right]$.
From the one-sided Fourier transform of $Z(t)$, we obtain the real and imaginary parts of $\hat{Z}(\omega)$ as
$\Real \hat{Z}(\omega) = D_{\infty} \e^{-|\omega\tau|}$ and
$\Imag \hat{Z}(\omega) = D_\infty [\e^{-\omega\tau}\Ei(\omega\tau) -\e^{\omega\tau}\Ei(-\omega\tau)]/\pi$,
being even and odd functions in $\omega$, respectively (\cref{fig:synthetic}a).
Here, $\Ei(\cdot)$ denotes the exponential integral, and
$D_\infty=\vth^2\tau \pi/2$ is the long-time limit of the diffusivity,
$D(t)=\int_0^t Z(s) \,\diff s = \vth^2 \tau \arctan(t/\tau)$.
In application of theorem~1.2 of ref.~\citenum{Mimica:JMAA2016}, we confirm that
\begin{equation}
  \lim_{\omega \to \infty} (-\omega)^{-1} \log\boldsymbol(\Real Z(\omega)\boldsymbol) =: \tau_\text{rc}
\end{equation}
yields the radius of convergence, $\tau_\text{rc} = \tau$, of the short-time expansion of $Z(t)$;
in particular, $\tau_\text{rc} > 0$.

Given $Z(\omega)$, the explicit expression for the complex memory kernel $\hat\gamma(\omega)$ and the friction $\zeta(\omega)$ follow from \cref{eq:def-memory,eq:def-friction}, see \cref{fig:synthetic}b. The low- and high-frequency asymptotes correspond to $\zeta(\omega \to 0) \simeq \zeta_0 (1+|\omega \tau|)$ and $\zeta(\omega \to \infty) \simeq \zeta_0 (\pi \omega\tau/2)^2 \, \e^{-|\omega \tau|}$, respectively, with $\zeta_0 = \kB T /D_{\infty}$.
The friction attains its maximum $\approx 1.2\zeta_0$ near $\omega_\mathrm{max} \approx 0.892 \tau^{-1}$
and falls off rapidly for larger $\omega$; the position of the maximum of $\Imag \hat\gamma(\omega)$ sets the onset frequency $\omega_c \approx 4.01 \tau^{-1}$.
The memory function $\gamma(t)$ is obtained numerically from $\zeta(\omega)$ using \cref{eq:backtransform},
with the short- and long-time asymptotes
$\gamma(t \to 0) \simeq 2 \tau^{-2} [1-5 (t/\tau)^2]$
and $\gamma(t \to \infty) \simeq -(\zeta_0/m)^2 (t/\tau)^{-2}$, respectively (\cref{fig:synthetic}c).

\clearpage
\subparagraph{Adapted Filon algorithm.}
The computation of the frequency-dependent friction requires a robust numerical Fourier transform, for which we developed a physics-enriched version of Filon's quadrature formula. 
The goal is to evaluate $\hat f(\omega) = \int_0^\infty \e^{\i \omega t} f(t) \,\diff t$ for a function $f(t)$ sparsely sampled on an irregular grid $t_0=0, t_1, \dots, t_n$ for an arbitrary set of frequencies $(\omega_j)$.
The idea of Filon's algorithm is to interpolate $f(t)$ by elementary functions between the grid points (usually parabolas), thereby reducing the Fourier integral to a finite sum of integrals, for which analytic expressions exist.
Anticipating that the normal physical decay of correlation functions is exponential, we approximate $f(t) \approx a_k \e^{-\gamma_k t}$ in the interval $[t_k, t_{k+1}]$ with $a_k$ and $\gamma_k$ fixed by $f(t_k)$ and $f(t_{k+1})$.
Then,
\begin{multline}
  \hat f(\omega) \approx \int_{0}^{t_1} \e^{\i \omega t} f(t) \,\diff t
    + \sum_{k=1}^{n-1} \int_{t_k}^{t_{k+1}} a_k \e^{(\i \omega - \gamma_k)t} \,\diff t \\
    + \int_{t_n}^\infty a_n \e^{(\i \omega -\gamma_n) t} \,\diff t \,.
  \label{eq:filon}
\end{multline}
Spurious low-frequency oscillations of the transform are removed by smoothly truncating the integral at $t_n$, here by
assuming a terminal exponential decay of $f(t)$, which leads to the last term on the r.h.s. of \cref{eq:filon}.
In order to preserve the short-time properties of $f(t)$ we fit a polynomial in $t^2$ to the first few data points and solve the integral analytically; this improves the high-frequency behaviour of $\hat f(\omega)$.

The dynamic friction $\zeta(\omega)$ and the memory function $\gamma(t)$ are obtained from MSD data as follows (\cref{fig:flowchart}):
The timescale-dependent diffusion coefficient, $D(t) := \partial_t \MSD(t) / 6$, is obtained from numerical differentiation. In all cases studied, it grows out from zero, has a maximum, and converges slowly towards the long-time diffusion constant $D_\infty = D(t \to \infty)$, see supplemental fig.~S1.
Using the above algorithm, we compute \citeM{Franosch:JNCS2011}
$\hat{Z}(\omega)=D_\infty-\i \omega \int_0^{\infty} \diff t \,\mathrm{e}^{\i \omega t}[D(t)-D_\infty]$.
Then, $\zeta(\omega)$ is given by \cref{eq:def-memory} and is transformed back to the time domain with the same algorithm [\cref{eq:backtransform}]; in particular, we use again a smooth, exponential cutoff.
In \cref{fig:synthetic}, the numerical procedure is successfully tested against the analytical model with high accuracy.

\subparagraph{Deconvolution in time domain.}

Inversion of the convolution in \cref{eq:VACF} yields the memory function $\gamma(t)$ directly \citeM{Berne:JCP1990}, without resorting to the frequency domain.
Numerically, it is not easy to obtain accurate and robust results, and a variety of algorithms have been developed, see ref.~\citenum{Kowalik:PRE2019} for a comparative study.
The presence of $\dot Z(t)$ in \cref{eq:VACF} is removed by integration, yielding
$Z(t)=Z(0)-\int_0^t \diff s \, G(t-s) Z(s)$
with the integrated memory $G(t):=\int_0^t \diff s \, \gamma(s)$.
Discretising on a uniform time grid, $t_i = i \Delta t$ ($i=0, 1, \dots$), and employing the trapezoidal rule for the integral, a recursion relation for $G_i:=G(t_i)$ with the initial value $G_0=0$ follows \cite{Kowalik:PRE2019}:
\begin{equation}
  G_i = \frac{1 - Z_i/Z_0}{\Delta t/2} - 2 \sum_{j=1}^{i-1} G_j Z_{i-j} / Z_0 \,, \quad i \geq 1 \,.
  \label{eq:Gi-VACF}
\end{equation}
Going beyond ref.~\citenum{Kowalik:PRE2019}, we introduce a predictor--corrector scheme to stabilise the numerical solutions:
In the predictor step, one evaluates $G_{i}^{*}$ and $G_{i+1}^{*}$ from \cref{eq:Gi-VACF}. Afterwards, the weighted average $G_i:=(G_{i-1}+3G_{i}^{*} + G_{i+1}^{*})/5$ manifests itself as the corrector step.
Results for $G(t)$ can be found in the supplemental fig.~S2.
Finally, the memory function $\gamma(t)=\partial_t G(t)$ is obtained by central differences.
If one starts from MSD data on a sparse (e.g., geometrically spaced) time grid, a cubic spline interpolation of the MSD in the variable $t^2$ is suitable to sample $Z(t)$ on a uniform grid of up to \SI{e5} points.

\subparagraph{Molecular dynamics simulations.}

Simulations of liquid water were performed with the GROMACS 5.1 package \citeM{Hess:JCTC2008}
using the SPC/E water model, which was shown to accurately reproduce the
linear absorption spectra of water from experiments and ab-initio MD simulations up to frequencies of about \SI{30}{THz} \citeM{Carlson:2020}.
The system of 3{,}007 molecules in a cubic box of size \SI{4.49}{nm} was equilibrated at \SI{300}{K} and \SI{1}{bar} following standard procedures \cite{Kowalik:PRE2019}.
Correlation functions were obtained from an NVE simulation over \SI{275}{ps} with the velocity-Verlet integrator and a time step of \SI{1}{fs}, using double floating-point precision to achieve good energy conservation.
The frequency-dependent viscosity was computed from additional NVE runs, totalling to \SI{49}{ns}.

For the other two liquids, we used the massively parallel software \emph{HAL's MD package}\citeM{Colberg:CPC2011,*HALMD} (version 1.0-$\alpha6$), permitting the sampling of dynamic correlations on a sparse time grid and featuring smoothly truncated potentials to virtually eliminate any energy drift.
The mono-atomic fluid consists of \SI{e5} particles interacting pairwise via the LJ potential,
$U(r)=4\epsilon\bigl[(r/\sigma)^{-12} - (r/\sigma)^{-6}\bigr]$, truncated for $r \geq 2.5\sigma$;
a unit of time is defined by $\tau_\mathrm{LJ} := \sqrt{m\sigma^2/\epsilon}$.
Equilibration in the NVE ensemble at number density $\rho=0.8 \sigma^3$ and thermal energy $k_\textrm{B}T =1.3 \epsilon$ followed the protocol in ref.~\citeM{Roy:JCP2016}.
The supercooled liquid was realised by a Kob--Andersen 80:20 binary mixture \citeM{Kob:PRL1994} of 64,000 LJ beads
at $\rho=1.2\sigma^{-3}$ and $T^*:= k_\textrm{B}T/\epsilon =0.6$, equilibrated over a time span of $9{,}000\,\tau_\mathrm{LJ}$, and we traced the species of the larger particles.
The chosen temperature is well below the melting point\citeM{Pedersen:PRL2018}, $T^*\approx 1.03$,
and is on the onset of universal scaling behaviour according to mode-coupling theory \cite{Goetze:MCT};
here, the value of the critical temperature is $T^*_\mathrm{MCT} \approx 0.43$.

The simulations generate trajectories $\vec r(t)$ of an ensemble of labelled particles for each fluid;
our main observable is the mean-square displacement $\MSD(t) := \expect{|\vec r(t) - \vec r(0)|^2}$ for lag time $t$.
For both liquids, single-particle MSDs were averaged from 10 production runs, each over \SI{e8} integration steps of length $0.001\,\tau_\mathrm{LJ}$.

Control simulations of a single particle in its pinned cage are based on an equilibrated sample of the LJ fluid with \SI{e6} particles. MSDs were recorded after equilibration of the mobile particle in its static environment over $100\,\tau_\mathrm{LJ}$ and were averaged over \SI{e6} different cages, computed in parallel, to remove spurious oscillations.
Technically, the setup was realised by making two initially identical copies of the fluid interact with each other: the first copy contains the immobile matrix, the second one the tracers (not interacting with each other).

\subparagraph*{Acknowledgements.}

FH is indebted to Thomas Franosch for introducing him to the field of complex transport.
We benefited from discussions with Lydéric Bocquet, Matthias Fuchs, Julian Kappler, and Klaus Kroy.
This research has been funded by Deutsche Forschungsgemeinschaft (DFG, German Research Foundation) through the grant SFB~1114 (projects B03 and C01) and under Germany's Excellence Strategy -- MATH+ : The Berlin Mathematics Research Center (EXC-2046/1) -- project ID: 390685689 (subproject EF4-4).
Further funding by the European Research Council (ERC Advanced H 2020 Grant ``NoMaMemo'') is
gratefully acknowledged.
Some of the data were produced with the supercomputer ``Lise'' (HLRN-IV) of the North-German Supercomputing Alliance.

\subparagraph*{Author contributions.}

AS, RN, and FH conceived the project and wrote the manuscript. BK and FH performed the simulations, and AS and FH analysed the data. AS carried out the analytical work.
All authors discussed the results and implications and commented on the manuscript at all stages.

\subparagraph*{Data availability.}

The datasets generated and analysed during the current study are available from the corresponding author upon reasonable request.

\subparagraph*{Code availability.}

Primary data were generated with open source software as indicated in the Methods section. The source code used to analyse the data for the current study is available from the corresponding author upon reasonable request.

\bibliographystyleM{aipnum4-1}
\bibliographyM{dynamic-friction}

\end{document}